\documentstyle [12pt] {article}

\newcommand{\gtwid}{\mathrel{\raise.3ex\hbox{$>$\kern-.75em\lower1ex
\hbox{$\sim$}}}}
\newcommand{\ltwid}{\mathrel{\raise.3ex\hbox{$<$\kern-.75em\lower1ex
\hbox{$\sim$}}}}
\newcommand{\beq}{\begin{equation}}
\newcommand{\eeq}{\end{equation}}
\newcommand{\beqs}{\begin{eqnarray}}
\newcommand{\eeqs}{\end{eqnarray}}

\catcode`@=11
\@addtoreset{equation}{section}
\@addtoreset{equation}{subsection}
\def\theequation{\ifnum\value{section}=0 \arabic{equation}\ignorespaces
\else \ifnum\value{section}=-1 A.\arabic{equation}\ignorespaces
\else \ifnum\value{subsection}=0 \thesection.\arabic{equation}\ignorespaces
\else \thesection.\arabic{subsection}.\arabic{equation}\ignorespaces
                           \fi
                      \fi
                 \fi}
\catcode`@=12

\begin{document}

\def\thefootnote{\fnsymbol{footnote}}
\baselineskip 6.0mm

\begin{flushright}
\begin{tabular}{l}
ITP-SB-95-1    \\
February, 1995
\end{tabular}
\end{flushright}

\vspace{8mm}
\begin{center}

{\Large \bf Complex-Temperature Properties of the Ising Model }

\vspace{3mm}

{\Large \bf on 2D Heteropolygonal Lattices}

\vspace{16mm}

\setcounter{footnote}{0}
Victor Matveev\footnote{email: vmatveev@max.physics.sunysb.edu}
\setcounter{footnote}{6}
and Robert Shrock\footnote{email: shrock@max.physics.sunysb.edu}

\vspace{6mm}
Institute for Theoretical Physics  \\
State University of New York       \\
Stony Brook, N. Y. 11794-3840  \\

\vspace{16mm}

{\bf Abstract}
\end{center}
Using exact results, we determine the complex-temperature phase diagrams of
the 2D Ising model on three regular heteropolygonal lattices,
$(3 \cdot 6 \cdot 3 \cdot 6)$ (kagom\'{e}), $(3 \cdot 12^2)$, and
$(4 \cdot 8^2)$ (bathroom tile), where the notation denotes the regular
$n$-sided polygons adjacent to each vertex.  We also work out the exact
complex-temperature singularities of the spontaneous magnetisation.
A comparison with the properties on the square, triangular, and hexagonal
lattices is given.   In particular, we find the first case where, even for
isotropic spin-spin exchange couplings, the nontrivial non-analyticities of
the free energy of the Ising model lie in a two-dimensional, rather than
one-dimensional, algebraic variety in the $z=e^{-2K}$ plane.
\vspace{16mm}

\pagestyle{empty}
\newpage

\pagestyle{plain}
\pagenumbering{arabic}
\renewcommand{\thefootnote}{\arabic{footnote}}
\setcounter{footnote}{0}

\section{Introduction}
\label{intro}

    The Ising model has long served as a prototype of a statistical mechanical
system which undergoes a phase transition with associated spontaneous symmetry
breaking and long range order.  The Ising model on a dimension $d=2$ lattice
has the great appeal that (for the spin 1/2 case with nearest-neighbor
interactions) it is exactly solvable; for the square lattice,
in the absence of an external magnetic field $H$, the free energy
was first calculated by Onsager \cite{ons}, and a closed-form expression for
the spontaneous magnetisation was first derived by Yang \cite{yang}.
Later, the zero-field model was also solved for the triangular and hexagonal
(= honeycomb) lattices; for a comparative review of the solutions on these
three lattices, see Ref. \cite{domb1}.  However, in addition to the square,
triangular, and hexagonal lattices, which involve tilings of the plane by one
type of regular polygon, there are also other 2D lattices which have the
property that all vertices are equivalent and all links (bonds) are of equal
length, but are comprised of tilings by
more than one type of regular polygons.  We shall denote the regular 2D
lattices which are comprised of tilings by one type of polygon as
``homopolygonal'' and those involving tilings by more than one type of polygon
as ``heteropolygonal''.  Indeed, as will be discussed further below, the
(zero-field) free energy and spontaneous magnetisation for the Ising model have
been calculated for three heteropolygonal 2D lattices.  The results yield
valuable insights into how the properties of the solution depend on the tiling
of the plane and, in particular, how they change when this tiling involves more
than one kind of polygon.

In the present paper, we
shall determine the complex-temperature phase diagrams and singularities of the
spontaneous magnetisation for the three particular heteropolygonal lattices for
which the Ising model has been solved.
There are several reasons for studying the properties of statistical
mechanical systems such as spin models with the
temperature variable generalised to take on complex values.  First, one can
understand more deeply the behaviour of various
thermodynamic quantities by seeing how they behave as analytic functions of
complex temperature.  Second, one can see how the physical phases of a given
model generalise to regions in appropriate complex-temperature variables.
Third, a knowledge of the complex-temperature singularities of quantities
which have not been calculated exactly helps in the search for exact,
closed-form expressions for these quantities.  This applies, in particular,
to the susceptibility of the 2D Ising model.  Complex-temperature properties
of the Ising model (zeros of the partition function and singularities of
thermodynamic and response functions) have been explored in a number of
papers for homopolygonal 2D lattices and for various 3D lattices
\cite{fisher, g69, dg, g75, ipz, ms, egj, chisq, chitri, chihc, ih}.
(Refs. \cite{ipz} and part of Refs. \cite{g69, dg} deal with the $d=3$ Ising
model.)  However, to our knowledge, analogous investigations of
complex-temperature properties have not been reported for heteropolygonal 2D
lattices.

   To begin, we need to describe the heteropolygonal lattices.
We shall use the standard mathematical notation for general lattices
\cite{gs}.  An important virtue of this notation is that the mathematical
symbol explicitly contains sufficient information for one to reconstruct the
lattice. A regular tiling of the plane is defined as one involving one or
more regular polygons, i.e. polygons each of whose sides are of equal
length. An Archimedean lattice is defined as a regular tiling of the plane
in which all vertices are
equivalent.  Thus each vertex has the same coordination number, which we shall
denote as $q$.  From the definition, it follows immediately that all of the
links on an Archimedean lattice are of equal length.  For this reason, it is
natural to consider the case of equal spin-spin couplings, and we shall do
this here. However, it should be noted that the links on an Archimedean lattice
are not, in general, all equivalent.  For the three homopolygonal Archimedean
lattices, it is obvious that all links are equivalent, but for the
heteropolygonal lattices, this equivalence only holds if (a) the lattice
consists of only two types of polygons and (b) each edge of the first type of
polygon is also an edge of the second type of polygon.  Among the eight
heteropolygonal Archimedean lattices, this condition is met only for the $(3
\cdot 6 \cdot 3 \cdot 6)$ (kagom\'{e}) lattice (see below).
The mathematical definition of, and notation for, an Archimedean lattice
$\Lambda$ are specified by
\beq
\Lambda = (\prod_{j=1}^q p_j)
\label{lambda}
\eeq
where $p$ denotes a regular $p$-gon.  The meaning of eq. (\ref{lambda}) is
that as one makes a small circuit around any vertex, one
traverses first the polygon $p_1$, next $p_2$, and so forth, finally traversing
$p_q$.  Because the starting point is irrelevant, the symbol is invariant under
cyclic permutations:
\beq
(p_1 \cdot p_2 \cdot \cdot \cdot p_q) =
(p_2 \cdot p_3 \cdot \cdot \cdot p_q \cdot p_1)
\label{cyclic}
\eeq
and so forth for other cyclic permutations.
If a $p$-sided polygon occurs $\ell$ times sequentially in
this product, one signifies this by $p^\ell$.  There are 11 such ($d=2$)
Archimedean lattices \cite{gs}.  Of the 11 Archimedean lattices, three are
homopolygonal.  These are the well-known square, triangular, and hexagonal
lattices, which are denoted, respectively, $(4^4)$, $(3^6)$, and $(6^3)$.
Note
that the dual of a homopolygonal lattices $(p^\ell)$ is $(\ell^p)$, so that it
is self-dual if and only if $p=\ell$, which occurs only for $p=4$.  The other
eight Archimedean lattices are heteropolygonal. The full set of these lattices
is listed in Table 1.
(One lattice, $(3^4 \cdot 6)$, occurs in two enantiomorphic (chiral) forms,
which are considered together in this counting.)
\begin{table}
\begin{center}
\begin{tabular}{|c|c|c|c|c|} \hline \hline & & & & \\
lattice & name & $q$ & bip. & sym. \\
& & & & \\ \hline \hline
$(3^6)$ & triangular & 6 & N & $z \to -z$ \\ \hline
$(4^4)$ & square     & 4 & Y & $z \to -z$, $z \to 1/z$ \\ \hline
$(6^3)$ & hexagonal  & 3 & Y & $z \to 1/z$ \\ \hline \hline
$(3 \cdot 6 \cdot 3 \cdot 6)$ & kagom\'{e} & 4 & N & $z \to -z$ \\ \hline
$(3 \cdot 12^2)$ & $-$ & 3 & N & n \\ \hline
$(4 \cdot 8^2)$ & bathroom tile & 3 & Y & $z \to 1/z$ \\ \hline
$(3^4 \cdot 6)$ & $-$ & 5 & N & n \\ \hline
$(3^3 \cdot 4^2)$ & $-$ & 5 & N & n \\ \hline
$(3^2 \cdot 4 \cdot 3 \cdot 4)$ & $-$ & 5 & N & n \\ \hline
$(3 \cdot 4 \cdot 6 \cdot 4)$ & $-$ & 4 & N & $z \to -z$ \\ \hline
$(4 \cdot 6 \cdot 12)$ & $-$ & 3 & Y & $z \to 1/z$ \\ \hline
\hline
\end{tabular}
\end{center}
\caption{Table of the 11 Archimedean lattices. The first three are the
well-known homopolygonal ones, and the remaining eight are heteropolygonal.
The entries in the column
denoted bip. indicate whether (Y,N) the lattice is bipartite. The entries in
the column denoted ``sym.'' indicate whether the complex-temperature phase
diagram and zeros of the partition function for the (spin 1/2, zero-field)
Ising model have the symmetries $z \to -z$ and/or $z \to 1/z$, or neither,
denoted by ``n''.}
\label{table1}
\end{table}

    The three heteropolygonal Archimedean lattices for which we shall
determine the complex-temperature phase diagrams and singularities of the
magnetisation are those for which the Ising model has been solved exactly,
namely the (i) $(3 \cdot 6 \cdot 3 \cdot 6)$, (ii) $(3 \cdot 12^2)$, and
(iii) $(4 \cdot 8^2)$ lattices.  For the reader's convenience, these lattices
are shown in Figs. 1-3, respectively. The $(3 \cdot 6 \cdot 3 \cdot 6)$
lattice is often called the
kagom\'{e} lattice, after a Japanese word for a similar type of basket
weave. The $(4 \cdot 8^2)$ lattice is often called the bathroom tile lattice.
In the literature, the $(3 \cdot 12^2)$ and $(4 \cdot 8^2)$ lattices are
frequently denoted by the shorthand symbols $3-12$ and $4-8$.   As was
observed by Utiyama \cite{utiyama}, one way to construct these lattices is to
start from a checkerboard lattice, replace each ``black'' square by a square
with $n_v$ additional vertical bonds, and then take certain of the spin-spin
couplings either to 0 or $\infty$.  One chooses $n_v=1$ to construct the
$(3 \cdot 6 \cdot 3 \cdot 6)$ and $(4 \cdot 8^2)$ lattices, and $n_v=3$ to
construct the $(3 \cdot 12^2)$ lattice.  Using this observation, Utiyama
discussed ingredients for an exact solution for the (zero-field) free energy
for these lattices \cite{utiyama}.  These lattices can also be obtained by
(possibly iterated) decoration and/or star-triangle operations from the three
homopolygonal lattices \cite{syozi72}.

   We recall that abstractly, the duality mapping for a $d$-dimensional
lattice maps $k$-cells to $d-k$ cells.  In our present case, with $d=2$,
this mapping interchanges 0-cells (vertices) and 2-cells (polygonal faces)
while taking 1-cells (links) to other links.
The duals of Archimedean lattices are Archimedean if and only if they are
homopolygonal; for the heteropolygonal Archimedean lattices, the dual lattices
are of a different type, called Laves lattices in the mathematical
literature \cite{gs,laves}.  A Laves lattice is defined as a regular
tiling of the plane involving a single (in general, non-regular) polygon. Thus,
the links of a Laves lattice are not, in general, of equal length,
and the vertices (sites) are not, in general, equivalent.
There is a one-to-one correspondence, namely that of duality, between the 11
Archimedean lattices and the 11 Laves lattices.  Clearly, the
number of different types of 0-cells (vertices) on the Laves lattice
$\Lambda^*$ which is the dual of the Archimedean lattice $\Lambda$ is equal
to the number of different 2-cells (polygons) on $\Lambda$.  Similarly, the
fact that the Laves lattice consists of only one type of (generally
non-regular) 2-cell (polygon) follows, by duality, from the fact that its dual
Archimedean lattice consists of only one type of 0-cell (vertex).
The standard mathematical notation for a Laves lattice $\Lambda$ is
\beq
\Lambda = [(\prod_{j=1}^p q_j)]
\label{lambdalave}
\eeq
the meaning of which is as follows: on a given polygon, as one makes a circuit
around its periphery, one passes first a vertex with coordination number $q_1$,
next a vertex with coordination number $q_2$, and so forth, finally coming to a
vertex with coordination number $q_p$, where $p$ denotes the number of
(generally non-equal) sides of the polygon, which is, of course, equal to the
number of vertices of this polygon.  The Laves lattice
$[(\prod_{j=1}^p q_j)]$ is the dual of the Archimedean lattice
$(\prod_{j=1}^q p_j)$ for $p=q$ and $p_j=q_j$, $j=1...p$.
As an example, the Laves lattice $[4 \cdot 8^2]$
(commonly called the union jack lattice) is shown in Fig. 4 and is evidently
the dual of the Archimedean lattice $(4 \cdot 8^2)$ (bathroom tile).
 From the duality relation with the Archimedean lattices, or directly from the
definiton, it follows that there are three Laves lattices for which the
vertices are, in fact, all equivalent; these are $[3^6]$, $[4^4]$, and
$[6^3]$. We shall denote these as homovertitial, and the remaining eight Laves
lattices with inequivalent vertices as heterovertitial, by analogy with the
homopolygonal and heteropolygonal Archimedean lattices.

   The solution for the (zero-field) free energy on each of the Archimedean
lattices immediately gives the solution on the corresponding dual Laves
lattice.  Thus, our determination of the complex-temperature phase diagrams
for the three heteropolygonal Archimedean lattices considered
here also yields a corresponding determination of the complex-temperature phase
diagrams for their dual Laves lattices.

\section{Model and Notation}
\label{model}

   Our notation is standard, so we
review it here only briefly.  We consider the spin $1/2$ Ising model on
the 2D lattices $\Lambda$ as discussed above at a temperature $T$ and
external magnetic field $H$ (where $H=0$ unless otherwise specified)
defined by the partition function
\beq
Z = \sum_{\{\sigma_n\}} e^{-\beta {\cal H}}
\label{zfun}
\eeq
with the Hamiltonian
\beq
{\cal H} = -J \sum_{<nn'>} \sigma_n \sigma_{n'} - H \sum_n \sigma_n
\label{ham}
\eeq
where $\sigma_n = \pm 1$ are the $Z_2$ spin variables on each site $n$ of the
lattice $\Lambda$, $\beta = (k_BT)^{-1}$, $J$ is the exchange constant,
$<n n'>$ denote nearest-neighbour sites,
and the units are defined such that the
magnetic moment which would multiply the $H\sum_n \sigma_n$ is unity.
(Hereafter, we shall use the term ``Ising model'' to denote the spin 1/2 Ising
model unless otherwise indicated.) We use the usual
notation $K = \beta J$, $h = \beta H$, $v = \tanh K$, $z = e^{-2K}$,
$u = z^2 = e^{-4K}$, and $\mu = e^{-2h}$. Note that $v$ and $z$ are related
by the bilinear conformal transformation
\beq
z = \frac{1-v}{1+v}
\label{bilinear}
\eeq
We record the symmetries
\beq
K \to -K \ \Rightarrow \ \{ v \to -v \ , \ \ z \to 1/z \ , \ \ u \to 1/u \}
\label{ksym}
\eeq
It will also be useful to introduce the common abbreviations
\beq
C \equiv \cosh(2K)
\label{cosh2k}
\eeq
\beq
S \equiv \sinh(2K)
\label{sinh2k}
\eeq
The reduced free energy per site is
$f = -\beta F = \lim_{N_s \to \infty} N_s^{-1} \ln Z$ in the thermodynamic
limit, where $N_s$ denotes the number of sites on the lattice.

\section{Some Basic Properties}
\label{theorems}

   We begin by discussing the phase boundaries of the model as a function of
complex temperature, i.e. the locus of points across which the free energy is
non-analytic.  The physical phases of the model include the phase where the
Z$_2$ symmetry is realised explicitly, viz., the paramagnetic (PM) phase,
and also, for dimension greater than the lower critical dimensionality (i.e.,
$d \ge 2$, for integer $d$), a phase where the $Z_2$ symmetry is spontaneously
broken with long-range ferromagnetic (FM) order, i.e. a nonzero spontaneous
(uniform) magnetisation, $M$.  For the lattices considered here which are
bipartite and hence involve no frustration for antiferromagnetic (AFM)
ordering, there is also a phase with AFM long range order, i.e. a nonzero
staggered magnetisation, $M_{st}$. As one can see from Table 1, of the three
heteropolygonal lattices considered here, only the $(4 \cdot 8^2)$
lattice is bipartite.  One defines the complex-temperature extensions of the
physical phases by analytic continuation in $K$ or an equivalent variable
such as $v$, $z$, or $u$.  There are, in general, also complex-temperature
phases which have no overlap with any physical phase.  As in our earlier work,
we label these as $O$ phases (where O denotes ``other''),
including subscripts
to distinguish them where there are several.  In cases where one O phase is
precisely the complex conjugate of another, we label the one with
$Im(z) \ge 0$ by O (with subscript where necessary) and its complex conjugate
by O$^*$.  (From eq. (\ref{bilinear}), it follows that in the $v$ plane, the
corresonding O and O$^*$ phases have $Im(v) \le 0$ and $Im(v) \ge 0$,
respectively.)

 As noted in Ref. \cite{ms}, there is an infinite periodicity
in complex $K$ under the shift $K \to K + n i \pi$, where $n$ is an integer,
and, for lattices
with even coordination number $q$, also the shift  $K \to (2n+1)i\pi/2$, as
a consequence of the fact that the spin-spin interaction
$\sigma_n\sigma_{n'}$ in ${\cal H}$ is an integer.  In particular, there is an
infinite repetition of phases as functions of complex $K$.  These repeated
phases are reduced to a single set by using the variables $v$, $z$ or $u$,
owing to the symmetry relation $K \to K+ n i \pi \Rightarrow \{ v \to v, \ \
z \to z, \ \ u \to u \}$.  It is thus convenient to use these variables here.

    As we have discussed in our earlier works \cite{chisq,chitri,chihc}, the
equations for the locus of points where the free energy is non-analytic also
serve to define the boundaries of the complex-temperature phases of the model.
Some of the loci of points where the free energy is non-analytic do not
actually separate any phases, but rather are arcs or line segments protruding
into various phases.  In passing we note that the free energy is, of course,
also trivially non-analytic at $K = \pm \infty$, i.e. $v = \pm 1$ or
$z = 0, \infty$.  This is obvious from eqs. (\ref{zfun}) and (\ref{ham}).
Because these points are isolated, they do not separate any
complex-temperature phases and hence will not be important here.

\vspace{2mm}

    Before proceeding, we list some theorems which formalise results in our
earlier work.  Although these are elementary, it will be useful to
record them here for our later discussion of the complex-temperature
phase diagrams.  Some of the theorems will be given in greater generality
than is needed here.

\vspace{2mm}

\begin{flushleft}

Theorem 1

\end{flushleft}

\vspace{2mm} Consider the (spin 1/2, zero-field) Ising model on an arbitrary
lattice. The loci of points in the complex $z$ and $v$ planes where the
free energy of the Ising model is non-analytic are symmetric under complex
conjugation, i.e. under reflection about the $Im(z)=0$ and $Im(v)=0$ axes,
respectively.  Hence the same is true of the complex-temperature phase
boundaries in these variables.  The same is also true of the zeros of the
partition function on finite lattices.

\vspace{2mm}

\begin{flushleft}

Proof

\end{flushleft}

One can prove the first two parts of this theorem either by starting with
finite lattices or by considering only the properties of the thermodynamic
limit.  If one uses finite lattices, this automatically also proves the last
part of the theorem, concerning zeros of the partition function.  The basic
property that one uses is that the partition function on a finite lattice is
a generalised polynomial.
(Here, a ``generalised polynomial'' in the variable $\zeta$ is defined
as function consisting of a finite sum of integral powers of $\zeta$, where
negative as well as positive powers are allowed.
Specifically, the partition function has the form
\beq
Z = \sum_{p=-N_\ell/2}^{N_\ell/2} c_p z^p
\label{zzh}
\eeq
where $N_\ell$ denotes the number of links on the lattice and
the $c_p$ are real.  (In fact, $c_p$ are not just real, but integral, and
also satisfy the symmetry $c_p = c_{N_\ell - p}$, but these properties will not
be needed here.)  The zeros of $Z$ are the zeros of the polynomial
\beq
P(z) = z^{N_\ell/2}Z = \sum_{p=0}^{N_\ell} c_p z^p
\label{pz}
\eeq
Since the polynomial $P(z)$ has real coefficients, it satisfies
\beq
P(z)^* = P(z^*)
\label{pconj}
\eeq
Hence, if $z_j$ is a zero of $P(z)$, then so is $z_j^*$. This proves the third
part of the theorem, on the zeros of the partition function for finite
lattices.  Taking the thermodynamic limit, one obtains the result that the
non-analyticities of the free energy and hence the complex-temperature phase
boundaries, are invariant under $z \to z^*$.  From the mapping (\ref{bilinear})
relating $z$ and $v$, the same result follows for the non-analyticities and
phase boundaries in the $v$ plane. \ $\Box$

    In passing, we note that if one uses periodic boundary conditions, then the
number of links is simply given by $N_\ell = (q/2)N_s$.  Moreover, for the part
of the theorem which deals with the free energy (in the thermodynamic limit),
an alternate way of seeing the result is to observe that $f$ has the form
\beq
f = \ln 2 + \int_{-\pi}^{\pi}\int_{-\pi}^{\pi} \frac{d\theta_1
d\theta_2}{(2\pi)^2} \ln \Bigl [ A + B(\theta_1,\theta_2) \Bigr ]
\label{fgeneral}
\eeq
where $A$ and $B$ are rational functions of $z$, or equivalently, $v$, with
real coefficients.  Aside from isolated points at $K=\pm \infty$ (i.e.,
$v=\pm 1$, \ $z=0$, $\infty$) and, for odd $q$, the point $z=-1$,
the above locus of points is the set where
$A+B(\theta_1,\theta_2)=0$. But since this amounts to a polynomial equation
in $z$ or $v$ with real coefficients, it follows that the roots of the
equation are either real or come in complex conjugate pairs.

   This theorem can be generalised in two ways.  Although we shall not need
either of these generalisations here, we state them for completeness.  First,
the theorem actually applies not just for the case of zero external field, but
also for the case of nonzero (real) magnetic field.  One can also show that it
is true for pure imaginary $h$.  Secondly, one can straightforwardly
generalise Theorem 1 to the case of the spin $s$ Ising model.
For this purpose, we define the normalisation of the spin values
to be $\sigma_n \in \{ -2s, -2s+2,...,2s-2, 2s\}$ in eqs. (\ref{zfun}) and
(\ref{ham}). Then the generalisation of
Theorem 1 reads as given there, without the restriction to spin 1/2.  For
integral $s$, a commonly used alternate normalisation is $\sigma_n \in \{-s,
-s+1, ..., s-1, s\}$; with this normalisation, the generalisation of Theorem 1
would use the modified definition $z=e^{-K}$.)

\begin{flushleft}

Theorem 2

\end{flushleft}

   For the (spin 1/2, zero-field) Ising model, if and only if the lattice has
even coordination number $q=2r$, the complex-temperature phase diagram is
invariant under the transformation $z \to -z$.  The same is true of the zeros
of the partition function for finite lattices.

\vspace{2mm}

\begin{flushleft}

Proof

\end{flushleft}

This follows by explicit calculation of the partition function for finite
lattices.  For odd $q$, $Z$ is a generalised polynomial in $z$, while for
even $q$, it is a generalised polynomial in $u=z^2$.
Using the definition of the free energy $f= \lim_{N_s \to \infty} N_s^{-1} \ln
Z$, it follows that the locus of points where the free energy is non-analytic
has the same symmetry.  $\Box$

\vspace{2mm}

Thus, for lattices with even $q$, a more compact way to display the
complex-temperature phase diagram is in the complex $u$ plane.

\vspace{2mm}

\begin{flushleft}

Theorem 3

\end{flushleft}

   For the (spin 1/2, zero-field) Ising model, if and only if the lattice is
bipartite, the complex-temperature phase diagram is invariant under the
transformation $z \to 1/z$ ($u \to 1/u$ if also $q$ is even).  The same is
true of the zeros of the partition function for finite lattices.

\vspace{2mm}

\begin{flushleft}

Proof

\end{flushleft}

  The statement that the lattice is bipartite means that it can be decomposed
into two sublattices, which we may denote as even ($e$) and odd ($o$), such
that each site on the even sublattice has as its nearest neighbors only sites
on the odd sublattice, and vice versa.  Since the spin-spin interaction in the
Hamiltonian only involves nearest-neighbor pairs, it follows that under the
mapping
\beqs
\sigma_e & \to \sigma_e \nonumber \\
\sigma_o & \to -\sigma_o \nonumber \\
J & \to -J
\label{evenodd}
\eeqs
the partition function $Z$ and hence also the free energy $f$ are invariant,
where $\sigma_e$ and $\sigma_o$ denote spins on the even and odd sublattices.
Hence
\beq
Z(K) = Z(-K) \ , \quad f(K) = f(-K)
\label{zfsym}
\eeq
The theorem follows, since $K \to -K$ is equivalent to $z \to 1/z$. \ $\Box$

\vspace{2mm}

\begin{flushleft}

Theorem 4

\end{flushleft}

In the $z$ plane, the zeros (and any divergences) of the spontaneous
magnetisation $M$ occur at either real values or at complex conjugate pairs of
values.

\vspace{2mm}

\begin{flushleft}

Proof

\end{flushleft}

Theorem 4 is a corollary of the generalisation of Theorem 1 to the case of
nonzero (real) external field $h$.  Since $M = \lim_{H \to 0} \partial
M/\partial h$, this generalisation of Theorem 1 implies that
\beq
M(z)^* = M(z^*)
\label{mzsym}
\eeq
Hence, in particular, the set of zeros of $M$ as a function of $z$ is invariant
under $z \to z^*$.  Similarly, the set of values of $z$ where $M$ diverges (if
this set is non-null) is invariant under $z \to z^*$.  \   $\Box$

\vspace{2mm}

    The following result is a corollary of this theorem:

\vspace{2mm}

\begin{flushleft}

Theorem 5

\end{flushleft}

On a bipartite lattice, in the $w = 1/z$ plane, the zeros (and any
divergences) of the spontaneous staggered magnetisation $M_{st}$ occur at
either real values or at complex conjugate pairs of values.

\vspace{2mm}

\begin{flushleft}

Proof

\end{flushleft}

This theorem follows from the fact that the symmetry (\ref{evenodd}) maps
the model with the ferromagnetic sign of the spin-spin coupling, $J>0$,
to the model with the antiferromagnetic sign, $J<0$.  This allows one
to obtain the staggered magnetisation $M_{st}$ immediately from the uniform
magnetisation;  in the physical AFM phase and, by analytic continuation,
thoughout the full complex-temperature extension of the AFM phase, $M_{st}$ is
given as
\beq
M_{st}(w) = M(z \to w)
\label{mstag}
\eeq
where $w = 1/z = e^{2K} = e^{-2|K|}$.  ($M_{st}$ vanishes identically
elsewhere.)  $\Box$

\vspace{3mm}

\begin{flushleft}

Theorem 6

\end{flushleft}

For lattices with odd coordination number $q$, the zero-field partition
function $Z$ of the (spin 1/2) Ising model vanishes at $z=-1$, and the free
energy contains a negatively divergent singularity at this point.

\vspace{2mm}

\begin{flushleft}

Proof

\end{flushleft}

 From the definition of the partition function and of $K$, we have
\beq
Z = \sum_{\{\sigma_n\}} \exp(K\sum_{<nn'>} \sigma_n \sigma_{n'})
\label{zk}
\eeq
Now, in general, for each link $<nn'>$, since $\sigma_n\sigma_{n'} = \pm 1$,
it follows that
\beq
e^{K\sigma_n\sigma_{n'}} = \cosh K + \sigma_n \sigma_{n'} \sinh K
\label{ecs}
\eeq
The $K$ values corresponding to $z=\lim_{\epsilon \to 0}(-1 \pm \epsilon)$
(where $\epsilon$ denotes a small real number) are
$K = -(1/2)\lim_{\epsilon \to 0} \ln(-1 \pm \epsilon) =
-(1/2)(\pm i \pi + 2n i \pi)$, where we
follow the usual convention of taking the branch cut for $\log z$ to lie
along the negative real $z$ axis, and $n$ indexes the Riemann sheet of the
logarithm.  To minimize unimportant minus signs, we consider
$z=\lim_{\epsilon \to 0}(-1 - \epsilon)$ and take the principal Riemann sheet
of the log, $n=0$; then $K=i\pi/2$.
Substituting this into eq. (\ref{ecs}), we have
\beq
e^{(i\pi/2)\sigma_n\sigma_{n'}} = i\sigma_n\sigma_{n'}
\label{es}
\eeq
so that
\beq
Z = \sum_{\{\sigma_n\}}\biggl (\prod_{<nn'>} i \sigma_n \sigma_{n'} \biggr )
\label{zexp1}
\eeq
Since the lattice has coordination number $q$, in this product over all links,
the $\sigma_n$ for each site $n$ appears $q$ times, and hence eq.
(\ref{zexp1}) can be rewritten as a product over all sites:
\beq
Z = i^{N_\ell}\sum_{\{\sigma_n\}}\Bigl ( \prod_n \sigma_n^q \Bigr )
\label{zexp2}
\eeq
Evaluating the summation over each spin, we have
\beq
Z = i^{N_\ell}\Bigl [ (+1)^q + (-1)^q \Bigr ]^{N_s}
\label{zexp3}
\eeq
Evidently, for odd $q$, the summation over $\sigma_n=\pm$ on each site
yields 0, and hence $Z=0$, so that the free energy is negatively
divergent at this point. \ $\Box$

\vspace{2mm}

   Among the lattices considered here, the hexagonal,
$(3 \cdot 12^2)$, and $(4 \cdot 8^2)$ ones have odd $q$ (in each case, $q=3$),
so that Theorem 6 implies that the respective free energy $f$ for each has a
divergent singularity at $z=-1$.

 From eq. (\ref{zexp3}) in the proof of Theorem 6, it is immediately evident
that for the (spin 1/2) Ising model on a lattice of even $q=2r$, the value of
the partition function on a finite lattice with periodic boundary conditions
(and hence the relation $N_\ell = (q/2)N_s$) is
\beq
Z(z=-1; q =2r) = i^{rN_s}2^{N_s}
\label{zm1qeven}
\eeq
while the (reduced) free energy satisfies
\beq
f(z=-1; q =2r) = r\ln i + \ln 2
\label{fzm1qeven}
\eeq
(independent of boundary conditions, since these do not affect the
thermodynamic
limit).  In eq. (\ref{fzm1qeven}), $\ln i = i\pi/2 + 2\pi i n$, where $n$
labels the Riemann sheet of the log.

\vspace{2mm}

   The points across which the free energy is non-analytic are related to the
zeros of the partition function.   A fundamental question concerns whether
these points (apart from the trivial ones at $K=\pm \infty$ and, for odd $q$,
at $z=-1$ as proved in Theorem 6)
lie on curves (including line segments) in the $z$ (or
equivalently, $v$) plane, or whether they lie in areas.  For homopolygonal
2D lattices with isotropic spin-spin exchange couplings, these points do lie on
curves.  (There is also numerical evidence for this in the case of 3D lattices
\cite{ipz}.)  It is also well known that for 2D homopolygonal lattices with
anisotropic couplings, these points lie, in general, in areas rather than on
curves \cite{areas}.   It is worthwhile to investigate this question for
heteropolygonal lattices, and we shall do so, as part of our general
determination of the complex-temperature phase diagrams.  Indeed, one of our
interesting results will be the finding that even for isotropic spin-spin
couplings, the zeros do not always lie on curves in the case of
heteropolygonal lattices.  We shall give a simple explanation of this
finding below.

   We now proceed with our analyses.

\section{$(3 \cdot 6 \cdot 3 \cdot 6)$ Lattice}
\label{lat3636}

   The (zero-field) free energy of the Ising model on the
 $(3 \cdot 6 \cdot 3 \cdot 6)$ (kagom\'{e}) lattice with equal spin-spin
couplings was first calculated explicitly by Kano and Naya \cite{kn}.  The
result is
\beq
f_{kag.}= \ln 2 + \frac{1}{6}\int_{-\pi}^{\pi}\int_{-\pi}^{\pi} \frac{d\theta_1
d\theta_2}{(2\pi)^2} \ln \Bigl \{ \frac{1}{4}  \Bigl [(C^3+S^3)^2 + 3C^2 -
2CS^2(C+S)P(\theta_1,\theta_2) \Bigr ] \Bigr \}
\label{fkag}
\eeq
where
\beq
P(\theta_1,\theta_2) = \cos(\theta_1) + \cos(\theta_2) +
\cos(\theta_1+\theta_2)
\label{ptheta}
\eeq
As was the case with the homopolygonal lattices, the connected locus of
points across which the free energy is non-analytic is the set of points for
which the logarithm in the integral of eq. (\ref{fkag}) vanishes.  These are
also the zeros of the partition function.  In terms of
the variable $u$, this vanishing condition is the equation
\beq
(21u^4+24u^3+18u^2+1) - 4u(1+u)(1-u)^2x = 0
\label{ueqkag}
\eeq
where $x$ represents $P(\theta_1,\theta_2)$ and takes values in the range
\beq
-\frac{3}{2} \le x \le 3
\label{xrange}
\eeq
 For $x=3$, there is a double real root at $u_{c,kag.}$, where
\beq
u_{c,kag.} = -1 + \frac{2}{\sqrt{3}} \simeq 0.15470054...
\label{uckag}
\eeq
and another double real root at $-1/(3u_{c,kag.} = -1 - 2/\sqrt{3}$.  As
indicated in the notation, $u_{c,kag.}$ is the physical critical point
separating the FM and PM phases.
As $x$ decreases from 3 in the range $-1 < x < 3$, each of these splits into
two pairs, which trace out the curves in Fig. 5(a).  For $x=-1$, these rejoin
in conjugate double roots at the multiple or intersection points $u_k$ and
$u_k^*$, where
\beq
u_k = \frac{1}{5}(-1+2i) = 5^{-1/2}e^{i\theta_k}
\label{uk}
\eeq
with
\beq
\theta_k = \pi - arctan(2) \simeq 116.57 ^ \circ
\label{thetak}
\eeq
As in our earlier works \cite{chisq,chitri,chihc}, we denote these as multiple
points, following the technical terminology of algebraic geometry, according to
which a multiple point of an algebraic curve is a point where two or more
branches (arcs) of the curve cross \cite{alg}.  We shall use the words
``intersection point'' and ``multiple point'' synonymously here.
The corresponding numerical values of $z_k=\pm 5^{-1/4}e^{i\theta_k/2}$ are
\beq
z_k = \pm (0.3515776 + 0.5688645i)
\label{zkvalue}
\eeq
 Finally, as $x$ decreases from $-1$ toward $-3/2$, these split again, with
one root moving away from $u_k$ upward along an arc of the circle defined by
\beq
|u- \frac{1}{3}| = \frac{2}{3}
\label{ucirclekag}
\eeq
toward the endpoint of this arc on the postive imaginary axis, given by
\beq
u_e = \frac{i}{\sqrt{3}}
\label{uekag}
\eeq
the conjugate root moving down from $u_k^*$ on the circle (\ref{ucirclekag})
toward $u_e^*$, and the other two roots moving along the circle
(\ref{ucirclekag}) toward the real axis.  Finally, at $x=-3/2$, two
roots occur at the endpoints $u_e$, $u_e^*$, and there is a double root at
the point $u=u_\ell \equiv -1/3$.

In terms of the commonly used variable $z$, the complex phase diagram is shown
in Fig. 5(b).  From the fact that the equation (\ref{ueqkag}) for the locus
of points where $f$ is non-analytic is symmetric under $z \to -z$, it follows
that the phase diagram in the $z$ plane has this symmetry also.
The physical critical point is $z_c = (-1+2/\sqrt{3})^{1/2}
\simeq 0.393320...$. The boundary of the complex-temperature FM phase crosses
the real $z$ axis at the points $\pm z_c$.  Corresponding to the intersection
point $u_k$ and its conjugate there are four intersection points $z_k =
5^{-1/4}e^{i\theta_k/2}$, $-z_k$, and $\pm z_k^*$.  The outermost points on the
boundaries of the O phases are given by $z_o = i|(-1 -2/\sqrt{3})|^{1/2}$ and
$z_o^*$.

  Finally, by a conformal mapping or directly from eq. (\ref{fkag}),
one can also determine the corresponding loci of points in the $v$ plane.  The
equation for this locus of points is
\beq
1-4v+10v^2-16v^3+22v^4-16v^5+10v^6-4v^7+v^8 -2v^2(1-v)^2(1+v^2)x=0
\label{veqkag}
\eeq
Up to an overall factor of $v^{-8}$, this equation is invariant under the
transformation $v \to 1/v$.  This together with the reality of the coefficients
implies that the locus of solutions is invariant under (i) $v \to v^*$ and (ii)
$v \to 1/v$.  This locus is shown in
Fig. 5(c).  The complex-temperature phases are marked on this figure and
consist of the PM and FM phases, together with two O phases which are related
to each other by complex conjugation.
 From the value of $u_c$ or $z_c$ and the relation (\ref{bilinear}),
it follows that the critical value of $v$ is
\beq
v_{c,kag.} = \frac{1}{2}(1+3^{1/2})\Bigl [ 1 - (2\sqrt{3}-3)^{1/2} \Bigr ]
\simeq 0.43542054...
\label{vckag}
\eeq
One sees that the kidney-shaped curve enclosing the FM and two O phases crosses
the real-$v$ axis at the points $v_c$ and $1/v_c \simeq 2.29663...$. The four
intersection points of the kidney-shaped curve and the arcs are given by $v_k =
(1-z_k)/(1+z_k)$, its complex conjugate, $v_k^*$, and their two reciprocals,
$1/v_k$ and $1/v_k^*$.  Similarly, the four endpoints of the arcs are given by
$v_e = (1-z_e)/(1+z_e)$, $v_e^*$, $1/v_e$, and $1/v_e^*$.  We find that the
inner endpoints of the arcs are the same distance from the origin as the
physical critical point, i.e. $|v_e| = v_c$.

    The phase structure consists of PM and FM phases, but, as is well known, no
AFM phase because of the frustration associated with AFM ordering.  These
characteristics are the same as those for the (isotropic) Ising model on
the triangular lattice.  In addition, Figs. 5(a) and 5(b) show two
complex-temperature phases denoted O and O$^*$ which have no overlap with any
physical phase.  In the $u$ variable (see Fig. 5(c)), these are mapped onto a
single O phase. In our previous work, we found O phases for the square and
triangular (but not hexagonal) lattices, in the $v$ and $z$ plane.  However,
for both the square and triangular lattices, each point $z$ in the O phase
was just the negative
of a corresponding point in the complex PM phase, and hence under the mapping
from the $z$ to $u$ plane, these were mapped to the same point, which could be
considered just the complex PM phase. We find for the kagom\'{e} lattice a new
property, viz., a distinct O phase which persists even in the $u$ plane.

The spontaneous magnetisation cannot be written in the same form
$M = (1 - (k_{<,\Lambda})^2)^{1/8}$ as for the regular unipolygonal lattices
$\Lambda = sq, \ tri, \ hc$.  Rather, it has the following form in the FM
phase (and vanishes identically elsewhere) \cite{naya54}
\beq
M_{kag.} = \frac{(1+3u)^{1/2}(1-u)^{1/2}}{(1+u)}\biggl (1 - (k_{<,kag.})^2
\biggr )^{1/8}
\label{mkagkl}
\eeq
where
\beq
k_{<,kag.} = \frac{2^{7/2}u^{3/2}(1+u)^{3/2}(1+3u^2)^{1/2}}
{(1-u)^3(1+3u)}
\label{klkag}
\eeq
i.e.,
\beq
M_{kag.} = \frac{(1-6u-3u^2)^{1/8}(1+2u+5u^2)^{3/8}(1+3u)^{1/4}}
{(1-u)^{1/4}(1+u)}
\label{mkag}
\eeq
These formulas may be analytically continued throught the complex-temperature
extension of the FM phase.
Note the factorizations $1-6u-3u^2 = (1-u/u_{c,kag.})(1+3u_{c,kag.}u)$ and
$1+2u+5u^2 = (1-u/u_k)(1-u/u_k^*)$.

   From eq. (\ref{mkag}) and our determination of the complex-temperature FM
phase in which the analytic continuation of this formula holds, it follows that
besides the well-known fact that $M_{kag.}$ vanishes continuously at the
physical critical point $u_c$, with exponent $\beta=1/8$, it also vanishes
continuously at the complex-temperature points $u_\ell=-1/3$ with exponent
\beq
\beta_{kag.,\ell} = \frac{1}{4}
\label{betaell}
\eeq
and at the points $u_k$ and $u_k^*$ with exponent
\beq
\beta_{kag.,k} = \frac{3}{8}
\label{betakagk}
\eeq
Elsewhere along the boundary of the complex-temperature FM phase, $M$ vanishes
discontinuously.
One may observe that the expression for $M$ has an apparent zero at
$u=-1/(3u_{kag.,c}) = -1 - 2/\sqrt{3} \simeq -2.1547$ (the point where the left
boundary in Fig. 5(a) crosses the real $u$ axis) and apparent
divergences at $u = 1$ and $u=-1$; however, none of these
singularities actually occurs since all of these points are outside of the
complex-temperature extension of the FM phase where the expression (\ref{mkag})
holds.  This is clear from the phase diagram in Fig. 5(a).

The characteristics of the points where $M$ vanishes continuously are listed
in Table 2, which
includes a comparison with the three 2D homopolygonal lattices.  The column
marked $z$ gives the symbol, if one was assigned, for each of the zeros (or
divergences) in the uniform (and, where relevant, staggered) magnetisation.
The column marked ``value'' lists either the analytic expression or, where this
is too long to fit into the space, a reference to the equation(s) where this is
given in the text.  The column marked adj. phases lists the phases which are
adjacent to the given point. For conciseness, some lines list a point and its
complex conjugate together; in these cases, the notation O$^{(*)}$ in the
adjacent phase column means O for the first point and O$^*$ for its complex
conjugate.  One may recall that the divergence in $M$ for the triangular
lattice occurs at $z=\pm 1/\sqrt{3}$ which are the endpoints of two line
segments protruding into the complex-temperature FM phase; this is indicated in
the table by the notation ``protrusion in FM''.
Other notation is explained in the table caption. It is also of
interest to perform this comparison for the lattices with even coordination
number, where all quantities can be expressed solely in terms of $u=z^2$; this
is done in Table 3.

\begin{table}
\begin{center}
\begin{tabular}{|c|c|c|c|c|c|c|} \hline \hline & & & & & & \\
$\Lambda$ & $z$ & value & $M$ & $M_{st}$ & $\beta$ & adj. phases \\
& & & & & & \\ \hline \hline
sq $(4^4)$
  & $z_c$ & $\sqrt{2}-1$ & zero & $-$ &  1/8 & FM-PM \\ \hline
  & $-z_c$ & $-(\sqrt{2}-1)$ & zero & $-$ &  1/8 & O-FM \\ \hline
  & $z_c^{-1}$ & $\sqrt{2}+1$ & $-$ & zero & 1/8 & PM-AFM \\
                                                                        \hline
  & $-z_c^{-1}$ & $-(\sqrt{2}+1)$ & $-$ & zero & 1/8 & AFM-O \\
                                                                        \hline
  & $z_s,z_s^*$ & $\pm i$ & zero & zero & 1/4 & O-FM-PM-AFM (mp)
                                                               \\ \hline\hline
tri $(3^6)$
  & $z_c$  & $1/\sqrt{3}$  & zero & N & 1/8 & FM-PM \\ \hline
  & $-z_c$ & $-1/\sqrt{3}$ & zero & N & 1/8 & O-FM \\ \hline
  & $z_s,z_s^*$ & $\pm i$  & zero & N & 3/8 & O-PM-FM (mp) \\
                                                                       \hline
  & $z_e,z_e^*$ & $\pm i/\sqrt{3}$ & div. & N & $-1/8$ & protrusion in FM
                                                                  \\
                                                                  \hline\hline
hex $(6^3)$
  & $z_c$ & $2-\sqrt{3}$ & zero & $-$ & 1/8 & FM-PM \\ \hline
  & $z_c^{-1}$ & $2+\sqrt{3}$ & $-$  & zero & 1/8 & PM-AFM \\ \hline
  & $z_s,z_s^*$ & $\pm i$     & zero & zero & 3/8 & FM-PM-AFM (mp)
                                                                    \\ \hline
  &    & $-1$         & div. & div.  & $-1/4$ & AFM-FM \\
                                                                  \hline\hline
$(3 \cdot 6 \cdot 3 \cdot 6)$
  & $z_c$ & $(-1+2/\sqrt{3})^{1/2}$ & zero & N & 1/8 & FM-PM \\
                                                                        \hline
  & $-z_c$ & $-(-1+2/\sqrt{3})^{1/2}$ & zero & N & 1/8 & PM-FM \\
                                                                      \hline
  &        & $\pm i/\sqrt{3}$  & zero & N & 1/4 & FM-O,
                                                           FM-O$^*$ \\ \hline
  & $\pm z_k$ & eq. (\ref{zkvalue}) & zero & N & 3/8
                                                    & FM-PM-O (mp) \\ \hline
  & $\pm z_k^*$ & eq. (\ref{zkvalue}) & zero & N & 3/8
                                          & FM-PM-O$^*$ (mp) \\ \hline \hline
$(3 \cdot 12^2)$
  & $z_c=z_{a+}$ & eqs. (\ref{zapm}),(\ref{zapf}) & zero & N & 1/8 & FM-PM \\
                                                                       \hline
  & $z_{a-}$ & eqs. (\ref{zapm}),(\ref{zamf}) & zero & N & 1/8 & PM-FM \\
                                                                       \hline
  &    & $\pm i/\sqrt{3}$ & zero & N & 1/4 & FM-O$^{(*)}$ \\
                                                                    \hline
  &    & $\pm i$  & zero & N & 3/8 & FM-PM-O$^{(*)}$ (mp) \\
                                                                     \hline
  & $z_r, z_r^*$ & $(1 \pm 2i)/5$ & zero & N & 3/8 &
                                                    FM-PM-O$^{(*)}$ (mp)
                                                          \\ \hline
  &   & $-1$   & div. & N & $-1/2$ & interior of FM \\
\hline\hline
$(4 \cdot 8^2)$
  & $z_c=z_{1+}$ & eqs. (\ref{z1pm}),(\ref{z1pnum}) & zero & $-$ & 1/8 &
                                                           FM-PM (mp;tn) \\
                                                                     \hline
  & $(z_{1-})^{-1}$ & eqs. (\ref{z1pminv}),(\ref{z1mnum}) & zero & $-$ & 1/8 &
                                               O$_1$-FM (mp;tn)\\ \hline
  & $z_{3-}^*, z_{3,-}$ & eqs. (\ref{z3pm}),(\ref{z3mnum}) & zero & $-$
& 1/8 & FM-O$_2^{(*)}$ (mp) \\ \hline
  & $z_c^{-1}=(z_{1+})^{-1}$ & eqs. (\ref{z1pminv}),(\ref{z1pnum}) & $-$
                                            & zero & 1/8 & PM-AFM (mp;tn)
                                                                \\ \hline
& $z_{1-}$ & eqs. (\ref{z1pm}),(\ref{z1mnum}) & $-$  & zero
                                    & 1/8 & AFM-O$_1$ (mp;tn) \\
                                                                   \hline
& $z_{3+}$, $z_{3+}^*$ & eqs. (\ref{z3pm}),(\ref{z3pnum}) & $-$ & zero
& 1/8 & AFM-O$_3^{(*)}$ (mp) \\
\hline \hline
\end{tabular}
\end{center}
\caption{Points at which $M$ and $M_{st}$ vanish continuously (or diverge) for
the $(3 \cdot 6 \cdot 3 \cdot 6)$ (kagom\'{e}), $(3 \cdot 12^2)$, and $(4 \cdot
8^2)$ (bathroom tile) lattices, together with the three homopolygonal 2D
lattices, for comparison.  Under the $M$ and $M_{st}$ columns, an
entry marked $-$ means that the point cannot be reached from within the FM and
AFM phases, respectively. The symbol N means that the model has no AFM phase.
 (mp) means a multiple (=intersection) point through two or more phase
boundary arcs pass. See text for further discussion.}
\label{table2}
\end{table}

\begin{table}
\begin{center}
\begin{tabular}{|c|c|c|c|c|c|c|c|} \hline \hline & & & & & & & \\
$\Lambda$ & $u$ & formula & value & $M$ & $M_{st}$ & $\beta$ & adj. phases \\
& & & & & & \\ \hline \hline
sq $(4^4)$
  & $u_c$ & $3-2\sqrt{2}$ & 0.172 & zero & $-$ &  1/8 & FM-PM \\ \hline
  & $u_c^{-1}$ & $3+2\sqrt{2}$ & 5.83 & $-$ & zero & 1/8 & PM-AFM \\ \hline
  & $u_s$ & $-1$ &   & zero & zero & 1/4 & AFM-FM-PM (mp) \\ \hline\hline
tri $(3^6)$
  & $u_c$  & 1/3    &  & zero & N & 1/8 & FM-PM \\ \hline
  & $u_s$ & $-1$    &  & zero & N & 3/8 & PM-FM (mp) \\ \hline
  & $u_e$ & $-1/3$  &  & div. & N & $-1/8$ &  protrusion in FM \\ \hline\hline
$(3 \cdot 6 \cdot 3 \cdot 6)$
  & $u_c$ & $-1+2/\sqrt{3}$ & 0.155 & zero & N & 1/8 & FM-PM \\ \hline
  & $u_\ell$  & $-1/3$  &  & zero & N & 1/4 & O-FM \\ \hline
  & $u_k,u_k^*$ & $(-1\pm 2i)/5$ & & zero & N & 3/8 & O-FM-PM (mp) \\
                                                       \hline \hline
\end{tabular}
\end{center}
\caption{Points in the $u$ plane at which $M$ and $M_{st}$ vanish
continuously (or diverge) for the lattices with even coordination number
$q$. Notation is the same as in Table 2.}
\label{table3}
\end{table}

\section{$(3 \cdot 12^2)$ Lattice}
\label{lat31212}

The (zero-field) free energy for the  $(3 \cdot 12^2)$ lattice with
isotropic couplings was given implicitly in \cite{syozi1,syozi72}.  For various
generalisations to unequal couplings, the free energy was given in
Refs. \cite{huck} and \cite{linchen}. Evaluating some intermediate expressions
and performing some algebra, we obtain the explicit expression, in terms of
$z$ (and $K=-(1/2)\ln z$),
\begin{displaymath}
f_{3-12} = \frac{3}{2}K + \frac{1}{2}\ln (1+z)
\end{displaymath}
\beq
+ \frac{1}{12}\int_{-\pi}^{\pi}\int_{-\pi}^{\pi} \frac{d\theta_1
d\theta_2}{(2\pi)^2} \ln \Bigl [ A_{3-12} + B_{3-12}P(\theta_1,\theta_2)
\Bigr ]
\label{f312}
\eeq
where
\beq
A_{3-12} = 1-6z+24z^2-42z^3+66z^4-42z^5+48z^6-6z^7+21z^8
\label{a312}
\eeq
\beq
B_{3-12} = -2z(1-z+2z^2)(1+z)(1-z)^4
\label{b312}
\eeq

Since this lattice has odd coordination number $q=3$, Theorem 6 states that, in
addition to the trivial singularities at $K=\pm \infty$, i.e.,
$z=0, \infty$, the free energy $f$ has a singularity at $z=-1$.  This is
evident in eq. (\ref{f312}).  If
one takes the branch cut for the logarithm to extend from $z=-1$ to
$z=-\infty$, then as one approaches $z=-1$ from the direction of the origin,
$f$ becomes negatively infinite.
In the case of the honeycomb lattice, as was noted in Ref. \cite{chihc}, this
point lies on the continuous locus of points where the argument of the
logarithm in the integrand analogous to (\ref{f312}) vanishes.  However, in
contrast to the honeycomb lattice, for the $(3 \cdot 12^2)$ and
$(4 \cdot 8^2)$ lattices, this singularity is an isolated
one.  Aside from these isolated singular points, the free energy also has
non-analyticities which arise from the vanishing of the argument of the
logarithm inside of the integral in eq. (\ref{f312}).
The condition for this vanishing is
\beq
A_{3-12}+B_{3-12}x=0
\label{zeq312}
\eeq
where $x$ represents $P(\theta_1,\theta_2)$ as given above in
eq. (\ref{ptheta}).  This yields the curve shown in Fig. 6(a).  As marked
there, these curves serve to bound the complex-temperature extensions of the FM
and PM phases and also two phases which have no overlap with any physical
phase, and are labelled O and O$^*$.  The border of the complex FM phase
crosses the real $z$ axis at the points
\beq
z_{a\pm} = \frac{1}{2}\Bigl [ -(\sqrt{3}+1) \pm
(4 + \frac{10}{3}\sqrt{3})^{1/2} \Bigr ]
\label{zapm}
\eeq
with numerical values
\beq
z_{a+} = 0.19710468...
\label{zapf}
\eeq
\beq
z_{a-} = -2.9291555...
\label{zamf}
\eeq
The physical critical point of the model is $z_{c,3-12} = z_{a+}$.
The intersection points nearest to the real $z$ axis are given by $z_r$ and
$z_r^*$, where
\beq
z_r = \frac{1-2i}{5} = 5^{-1/2}e^{i\theta_k}
\label{zr}
\eeq
where $\theta_k$ was given above in eq. (\ref{thetak}) (for certain formally
analogous points in the $u$ plane).  The insection points farther out from the
real $z$ axis occur at $z = \pm i$.
For $x=3$, there are four double roots at $z_{a+}$, $z_{a-}$, $z_b$, and
$z_b^*$, where
\beqs
z_b & = \frac{1}{2}\Bigl [ (\sqrt{3}-1) +
(4 - \frac{10}{3}\sqrt{3})^{1/2} \Bigr ] \nonumber \\
    & \nonumber \\
    & \simeq 0.36602540 + 0.66586461i
\label{zb}
\eeqs
As can be seen from Fig. 6(a), the points $z_b$ and $z_b^*$ lie roughly at the
centers of the outer boundaries of the respective O and O$^*$ phases.
As $x$ decreases from 3 in the range $-1 \le x \le 3$, these double roots split
apart into pairs the members of which move away from these four points, tracing
the curves shown in Fig. 6(a).  At $x=-1$, these points rejoin at the four
intersection points $z = \pm i$ and $z = z_r$ and $z_r^*$.  As $x$ decreases
further in the range $-3/2 \le x /\le -1$, these four double roots split again,
with two of the roots moving toward each other on the arc forming the inner
boundary of the O phase adjacent to the FM phase, and similarly with two
other roots on the inner boundary of the O$^*$ phase, while the other four
roots move outward away from the origin along the two arcs and their complex
conjugates.  Finally, at $x=-3/2$, two pairs of roots on the inner arcs join to
form double roots at $z = \pm i/\sqrt{3}$, while the four outer roots reach the
endpoints of the arcs.  These endpoints are as follows:
\beq
z_{e\pm} = \frac{1}{2}\biggl [ e^{i \pi/3} \pm \Bigl (e^{2 i \pi/3} -
(4/3)\sqrt{3}e^{i \pi/6} \Bigr )^{1/2} \biggr ]
\label{zepm}
\eeq
with the values
\beq
z_{e+} \simeq 0.29556791 - 0.3588689 i
\label{zepnum}
\eeq
\beq
z_{e-} \simeq 0.20443209 + 1.2248943 i
\label{zemnum}
\eeq
are, respectively, the endpoints of the upper parts of the arcs in the
upper and lower half $z$ plane.  Their complex conjugates $z_{e+}^*$ and
$z_{e-}^*$ form the endpoints of the lower parts of these arcs.

   From implicit expressions given in Ref. \cite{huck}, we can express
the spontaneous magnetisation as
\beq
M_{3-12} = \frac{(1-6z+6z^2-6z^3-3z^4)^{1/8}(1+3z^2)^{1/4}
(1+z^2)^{3/8}(1-2z+5z^2)^{3/8}}{(1-z)(1+z)^{1/2}(1-z+2z^2)}
\label{m312}
\eeq
By analytic continuation, this expression holds throughout the
complex-temperature extension of the FM phase, which we have determined, as
shown in Fig. 6(a).  Note the factorization
\beq
1-6z+6z^2-6z^3-3z^4 = -3(z-z_{a+})(z-z_{a-})(z-z_b)(z-z_b^*)
\label{fact312a}
\eeq
Moreover, similar to another formula above (for $u$ rather than $z$), we have
the factorization $1-2z+5z^2 = (1-z/z_r)(1-z/z_r^*)$.   Finally, we note that
$1-z+2z^2=2(z-z_o)(z-z_o^*)$, where
\beq
z_o = \frac{1}{4}(1 + i\sqrt{7}) = 0.25 +i 0.66143783...
\label{zo}
\eeq
These points $z_o$ and $z_o^*$ lie in the middle of the two complex conjugate O
phases.  Thus, $M_{3-12}$ vanishes continuously at the eight points $z_{a+}$,
$z_{a-}$, $\pm i/{\sqrt{3}}$, $\pm i$, $z_r$, and $z_r^*$.  In Table 2 we list
these zeros, together with the corresponding exponents $\beta$ and the phases
which are adjacent at each point.
Elsewhere along the boundary of the complex-temperature FM
phase, $M$ vanishes discontinuously.
$M_{3-12}$ also has a divergence at at
the isolated point $z=-1$ in the interior of the complex-temperature FM phase,
where the free energy itself is also singular.  In addition to these actual
singularities, the formal expression (\ref{m312}) has apparent zeros at the
points $z_{b+}$, $z_{b-}=z_{b+}^*$, $z_o$, and $z_o^*$,
but these are not zeros of the actual
magnetisation, since these points cannot be reached from within the
complex-temperature FM phase where the expression (\ref{m312}) holds.  As one
can see from Fig. 6(a), $z_b$ lies on the boundary separating the O and PM
phase, and similarly $z_b^*$ lies on the boundary between the O$^*$ and PM
phase. The points $z_o$ and $z_o^*$ lie in the interior of the O and O$^*$
phases, respectively.

\section{$(4 \cdot 8^2)$ Lattice}
\label{lat488}

    In contrast to the $(3 \cdot 6 \cdot 3 \cdot 6)$ and $(3 \cdot 12^2)$
lattices, the $(4 \cdot 8^2)$ (bathroom tile) lattice is bipartite, and the
physical phase diagram consists of a PM, FM, and also AFM phase.  One way to
obtain the (zero-field) free energy is from the free energy for the dual $[4
\cdot 8^2]$ (union jack) lattice.  An explicit calculation of the latter for
the case of equal spin-spin couplings was given in \cite{vaks}.  (See also the
remarks in Ref. \cite{utiyama}; for the union jack lattice with general
couplings, which we
shall not need here, see also Refs. \cite{baxterchoy} and \cite{wulin}.)
Using this duality relation, one easily obtains the free energy for the
$(4 \cdot 8^2)$ lattice, in terms of $z$ (and $K=-(1/2)\ln z$), as follows:
\begin{displaymath}
f_{4-8} = \frac{3}{2}K + \frac{1}{2}\ln(1+z)
\end{displaymath}
\beq
 + \frac{1}{8}\int_{-\pi}^{\pi}\int_{-\pi}^{\pi} \frac{d\theta_1
d\theta_2}{(2\pi)^2} \ln \Bigl [ A_{4-8}(z) + B_{1;4-8}(z)
(\cos\theta_1 + \cos\theta_2)
 + B_{2;4-8}(z)\cos\theta_1\cos\theta_2 \Bigr ]
\label{f48z}
\eeq
where
\beq
A_{4-8}(z) = (1+z^2)^2(1-4z+10z^2-4z^3+z^4)
\label{a48}
\eeq
\beq
B_{1;4-8}(z) = 2z(1-z)^3(1+z)(1+z^2)
\label{b148}
\eeq
\beq
B_{2;4-8}(z) = -4z^2(1-z)^4
\label{b248}
\eeq
In addition to the trivial singularities at $K=\pm \infty$, i.e.,
$z=0, \infty$, and the additional isolated singularity
at $z=-1$, the free energy has non-analyticities where the argument of the
logarithm in the integrand of eq. (\ref{f48z}) vanishes.  These are given by
the equation
\beq
A_{4-8}(z) + B_{1;4-8}(z)(\cos\theta_1 + \cos\theta_2) +
B_{2;4-8}(z)\cos\theta_1\cos\theta_2 = 0
\label{zeq48}
\eeq
We find that the locus of points which are solutions to this equation fall,
in general, into areas rather than curves, in the $z$ or, equivalently, the
$v$ planes.  These areas degenerate into points
at certain special locations.  In Fig. 7(a) we show the solution to
eq. (\ref{zeq48}) in the $z$ plane, for a grid of values of $\theta_1$ and
$\theta_2$.  Because of the finite grid of values of $\theta_1$ and
$\theta_2$ used to make the plots, the zeros show a striped structure in
certain regions; it is understood that in the limit where the aforementioned
grid of values of $\theta_1$ and $\theta_2$ becomes infinitely fine, these
stripes would merge into coherent areas; the boundaries of these areas are
easily inferred from the plot.
The phases include the complex-temperature extensions of the physical
FM, PM, and AFM phases, as marked, together with five phases which have no
overlap with any physical phases, labelled O$_1$, O$_2$, O$_2^*$, O$_3$, and
O$_3^*$.  The rest of the $z$ plane involves various continuous regions of
points where the free energy is non-analytic (forming the thermodynamic limits
of zeros of the partition functions for finite lattices).
We shall discuss the special locations where the areas of zeros of
the partition function degenerate into points below, in conjunction with an
analysis of the spontaneous uniform and staggered magnetisations.  The
corresponding diagram in the $v$ plane is shown in Fig. 7(b).  As in Fig. 7(a),
because of the finite grid of values of $\theta_1$ and $\theta_2$ used to make
the plot, the zeros exhibit a striped or dotted structure in certain
regions; again, it is understood that for a dense set of values of
$\theta_1$ and $\theta_2$, the zeros in these regions form a continuous set.

    To our knowledge, this finding constitutes the first known case of an Ising
model with isotropic spin-spin couplings, where the non-analyticities of the
free energy (aside from the trivial ones at $K = \pm \infty$ and, for odd $q$,
at $z=-1$, as proved in Theorem 6) form a two-dimensional, rather than
one-dimensional, algebraic variety, i.e., lie in areas rather than on curves
(including line segments). It is easy to understand the basis for this
result.  The locus of points at which the free energy is
non-analytic (aside from the above-mentioned trivial singularities)
is determined by the condition
that in the integral occuring in the free energy, the expression in the
argument of the logarithm vanishes.  For homopolygonal lattices, this
expression reduces to the equation
\beq
A_{\Lambda}(z) + B_{\Lambda}(z)x = 0
\label{logeq}
\eeq
for isotropic spin-spin couplings, where $x$ represents the function
$P(\theta_1,\theta_2)=\cos \theta_1+\cos\theta_2$ for the square lattice and
$P(\theta_1,\theta_2)= \cos \theta_1 +
\cos \theta_2 + \cos(\theta_1+\theta_2)$ for the triangular and hexagonal
(honeycomb) lattices.  (Thus for the square lattice, $-2 \le x \le 2$, while
for the triangular and hexagonal lattices, $-3/2 \le x \le 3$.)
The solutions to eq. (\ref{logeq}) form a one-dimensional algebraic variety,
specifically, algebraic curves (including possible line
segments).  In contrast, for unequal spin-spin exchange constants
$J_i$, where $i=1,2$ for the square lattice, and $i=1,2,3$ for the triangular
and hexagonal lattices, the condition for the argument of the logarithm in the
integrand to vanish is of the form
\beq
A_{sq} + B_{sq;1}\cos\theta_1 + B_{sq;2}\cos\theta_2 = 0
\label{logeqsqaniso}
\eeq
for the square lattice, and
\beq
A_{\Lambda} + B_{\Lambda;1}\cos\theta_1 + B_{\Lambda;2}\cos\theta_2
+ B_{\Lambda;3}\cos(\theta_1+\theta_2)=0
\label{logeqthcaniso}
\eeq
for the triangular and hexagonal lattices, where in each case, the various $A$
and $B$ functions depend on the $z_i$, with $z_i = e^{-2K_i}$ and $K_i = \beta
J_i$.  Now let the (anisotropic) ratios of the $J_i$'s be fixed.  Then
eqs. (\ref{logeqsqaniso}) and (\ref{logeqthcaniso}) depend on two
independent real (periodic) variables, $\theta_1$ and $\theta_2$.
It follows that the
solutions, in general, form a two-dimensional algebraic variety, specifically
areas (which may degenerate to points at special values of $z$).

   The heteropolygonal lattices $(3 \cdot 6 \cdot 3 \cdot 6)$ and $(3
\cdot 12^2)$ are similar to the homopolygonal lattices in this respect, i.e.,
if the spin-spin couplings on each link are equal, then the condition for the
argument of the logarithm in the integrand to vanish is of the form
(\ref{logeq}).  However, as is evident from eq. (\ref{f48z}), this is not the
case for the $(4 \cdot 8^2)$ lattice.  That is, even if the spin-spin
couplings are equal for each link, the condition for the vanishing of the
argument of the logarithm in eq. (\ref{f48z}) is of the form (\ref{zeq48}),
which depends on two independent real (periodic) variables
$\theta_1$ and $\theta_2$.

\vspace{2mm}

   The expression for the spontaneous magnetisation was conjectured by Lin et
al. \cite{linetal} and proved by Baxter and Choy \cite{baxterchoy}.  As with
the other heteropolygonal lattices $\Lambda$, $M$ can be written (where it is
nonzero) as a prefactor times $(1-(k_{<,\Lambda})^2)^{1/8}$.  Let us define
(using the subscript $4-8$ to denote $(4 \cdot 8^2)$)
\beq
k_{<,(4-8)} = \frac{8z^2(1-2z+4z^2-2z^3+z^4)}{(1-z)^4(1+z^2)^2}
\label{kl48}
\eeq
and
\beq
a = \frac{(1+z^2)}{(1+z)^{1/2}(1-z+3z^2+z^3)^{1/2}}
\label{afactor}
\eeq
Then $M_{4-8} = a\Bigl ( 1-(k_<)^2 \Bigr )^{1/8}$, i.e.,
\beq
M_{4-8} = \frac{(1-4z-z^4)^{1/8}(1+4z^3-z^4)^{1/8}(1+z^2)^{1/2}
(1-2z+6z^2-2z^3+z^4)^{1/4}}{(1-z)(1+z)^{1/2}(1-z+3z^2+z^3)^{1/2}}
\label{m48}
\eeq
within the physical FM phase and, by analytic continuation, throughout the
complex-temperature extension of the FM phase; $M_{4-8}$ vanishes identically
elsewhere. We observe first that under the transformation $z \to 1/z$, the
numerator of this expression goes into itself times a power of $z$.
It follows that the set of formal zeros of this numerator is invariant under
the mapping $z \to 1/z$.  In particular, under this mapping, the first and
second quartic expressions in the numerator of $M_{4-8}$ are interchanged,
up to an overall power of $z$: $z \to 1/z$ $\Rightarrow$ $(1-4z-z^4) \to
-z^{-4}(1+4z^3-z^4)$.  The roots of these two quartics are therefore
reciprocals of each other.  Under this mapping, the other two factors,
$(1+z^2)$ and $(1-2z+6z^2-2z^3+z^4)$ transform into themselves, up to overall
powers of $z$, and consequently, their zeros also come in reciprocal pairs
(since, furthermore they do not have either of the self-reciprocal numbers
$z=\pm 1$ as roots). We note the factorizations
\beq
1-4z-z^4 = \Bigl [1-\sqrt{2}(\sqrt{2}+1)z-(\sqrt{2}+1)z^2 \Bigr ]
\Bigl [1-\sqrt{2}(\sqrt{2}-1)z +(\sqrt{2}-1)z^2 \Bigr ]
\label{fac481}
\eeq
\beq
1+4z^3-z^4 = \Bigl [ 1+\sqrt{2}z-(\sqrt{2}-1)z^2 \Bigr ]
\Bigl [ 1-\sqrt{2}z+(\sqrt{2}+1)z^2 \Bigr ]
\label{fac482}
\eeq
\beq
1-2z+6z^2-2z^3+z^4 = \Bigl [1-2e^{i\pi/3}z+z^2 \Bigr ]
\Bigl [1-2e^{-i\pi/3}z+z^2 \Bigr ]
\label{fac483}
\eeq
The expression (\ref{fac481}) has zeros at
\beq
z_{1\pm} = -\frac{1}{2}\biggl [ \sqrt{2} \pm (-2 + 4\sqrt{2})^{1/2} \biggr ]
\label{z1pm}
\eeq
\beq
z_{2\pm} = \frac{1}{2}\biggl [ \sqrt{2} \pm i(2 + 4\sqrt{2})^{1/2} \biggr ]
\label{z2pm}
\eeq
while the expression (\ref{fac482}) has zeros at
\beq
(z_{1\pm})^{-1} =
\frac{1}{2}\biggl [ 2 + \sqrt{2} \pm (10 + 8\sqrt{2})^{1/2}\biggr ]
\label{z1pminv}
\eeq
\beq
(z_{2\mp})^{-1}=
\frac{1}{2}\biggl [2 - \sqrt{2} \pm (10 - 8\sqrt{2})^{1/2}\biggr ]
\label{z2pminv}
\eeq
The expression (\ref{fac483}) has zeros at $z_{3\pm}$ and $z_{3\pm}^*$, where
\beq
z_{3\pm} = e^{i\pi/3} \pm (e^{2i\pi/3}-1)^{1/2}
\label{z3pm}
\eeq
Note that $z_{3-}=z_{3+}^{-1}$. These roots have the approximate numerical
values
\beq
z_{1+}=0.2490384 \ , \quad (z_{1+})^{-1}=4.0154454
\label{z1pnum}
\eeq
\beq
z_{1-}=-1.663252 \ , \quad (z_{1-})^{-1}=-0.60123183
\label{z1mnum}
\eeq
\beq
z_{2\pm}=0.7071068 \pm 1.383551i
\label{z2pmnum}
\eeq
\beq
(z_{2\mp})^{-1} = 0.2928932 \pm 0.5730856i
\label{z2pminvnum}
\eeq
\beq
z_{3+}= 0.8406250 + 2.137255i
\label{z3pnum}
\eeq
\beq
z_{3-}= (z_{3+})^{-1} = 0.1593750 - 0.4052045i
\label{z3mnum}
\eeq
Only a subset of these formal zeros are zeros of the actual magnetisation,
namely those which can be reached from within the complex-temperature FM phase
(they lie on the border of this phase) where the formula (\ref{m48}) holds;
the other zeros are spurious, since they occur at points which cannot be
reached in this manner, where consequently the formula does not apply.  The
true zeros of $M$ are at $z_c = z_{1+}$, $1/z_{1-}$, $z_{3-}^*$, and
$z_{3-}$.  The first two of these are, respectively, the physical
critical point and the point at which the left boundary of the
complex-temperature FM phase crosses the real $z$ axis.  As one can see
from Fig. 7(a), $z_{3-}^*$ lies where certain areas of zeros of the partition
function degenerate to a single
point separating the complex-temperature FM and O$_2$ phases, and similarly for
the complex conjugate point $z_{3-}$ separating the FM and O$_2^*$ phases.
Indeed, these four points are precisely the full set of points on the border
of the complex-temperature FM phase where various areas of zeros degenerate
to single points.  The exponents with which $M_{4-8}$ vanishes continuously at
these zeros are all equal to $1/8$.  As one moves upward, away from the origin
along the $Im(z)$ (vertical) axis in Fig. 7(a), one encounters a boundary
followed by a dense set of points where the free energy is non-analytic
before one reaches the point $z=i$.  This boundary prevents analytic
continuation, so that the zero at $z=i$, and similarly, the zero at $z=-i$,
in the expression (\ref{m48}) are not true zeros of $M$.
Note that the points $\pm i$ lie where the O$_1$ and PM phases
are directly adjacent, and separated by these points.  Similarly, all of the
remaining zeros in the numerator of the expression (\ref{m48}) are not
actual zeros of $M$.  However,
it is of some interest to observe where these spurious zeros lie.
The point $z_{2+}$ is located where the border between the O$_3$ and PM
phases narrows to zero thickness (elsewhere this border is comprised of arcs
forming strips, as one can see from Fig. 7(a)).  An analogous comment applies,
{\it mutatis mutandis}, for the complex conjugate point $z_{2+}^*$ and the
border between the O$_3^*$ and PM phase.  The reciprocal point $1/z_{2-}$ is
located where the border between the O$_2$ and PM phase narrows to zero
thickness, and similarly for the point $1/z_{2+}^*$ and the border between the
O$_2^*$ and PM phases.
The point $z_{3+}$ lies where two areas of zeros of the
partition function degenerate to a point separating the O$_3$ and AFM phases,
and similarly for $z_{3}^*$ and the border between the O$_3^*$ and AFM phases.
Finally, none of the the apparent divergences in $M_{4-8}$ actually occurs;
$z=1$ lies in the interior of the PM phase;
$z=-1$ is in the interior of the O$_1$
phase, and among the roots of the cubic $1-z+3z^2+z^3$, $z \simeq -3.38$ is in
the AFM phase, while $z \simeq 0.191 \pm 0.509i$ lie in the interiors of the
O$_2$ and O$_2^*$ phases, respectively.  In the case of the point $z=-1$, it is
interesting to compare this with the situation for the $(3 \cdot 12^2)$
lattice.  In both cases, since $q$ is odd, $f$ is singular at $z=-1$.  For the
$(3 \cdot 12^2)$ lattice, this point lies within the complex-temperature
extension of the FM phase, and $M$ is also singular (divergent) at this point.
In contrast, for the $(4 \cdot 8^2)$ lattice, this point is not within the
complex-temperature FM phase, but rather in the interior of one of the O
phases, and hence there is no associated singularity in $M$ at this point.

   As with other bipartite lattices, the staggered spontaneous magnetisation
$M_{st,4-8}$ is given by eq. (\ref{mstag}) in terms of $M_{4-8}$ in the
complex-temperature extension of the AFM phase and vanishes elsewhere.
We may thus read off the zeros of $M_{st,4-8}$ immediately; these are precisely
the inverses of the four zeros of $M$, i.e., $z_c^{-1}$, $z_{1-}$,
$(z_{3-})^{-1}=z_{3+}$, and $(z_{3-}^*)^{-1}=z_{3+}^*$.
The first point is the physical critical point separating the PM and AFM
phase. The second is the point where the boundary between the AFM and O phases
crosses the negative real $z$ axis.  The third is where the border
between the O$_3$ and AFM phases narrows to zero thickness, and similarly for
the fourth point, separating the O$_3^*$ and AFM phases.
The actual zeros of $M$ and $M_{st}$ are listed in Table 2.

\vspace{2mm}

    In Table 4 we compare the general properties of the complex-temperature
phase diagrams for the Ising model on homopolygonal and heteropolygonal 2D
Archimedean
lattices, in terms of the variable $z$, or equivalently, $v$.
Since we have already indicated in Table 2 which lattices have AFM
phases, and since all have PM and FM phases, we do not include that information
here.  The entries in the column marked $N_O$ are the number of O phases. We
observe that the Ising model on heteropolygonal lattices exhibits more of these
O phases than on homopolygonal lattices.
As we have discussed in our earlier
papers \cite{chisq,chitri,chihc}, the points at which the curves (or line
segments, if present) cross are singular points of the algebraic curves,
in the technical terminology of algebraic geometry \cite{alg}.  This usage of
the term should,
of course, not be confused with the different meaning of ``singular point''
in statistical mechanics; for example, the point $z_c=(-1+2/\sqrt{3})^{1/2}$ of
the Ising model on the kagom\'{e} lattice is a singular point (actually, the
physical critical point separating the PM and FM phases) in the statistical
mechanical sense, but is not a
singular point of the curve forming the outer boundary of the
complex-temperature FM phase, in the algebraic geometry sense (see Fig.
5(b)).  A singular point of the algebraic curve is denoted as
a multiple point of index $n_b$ if $n_b$ branches (arcs) of the curve pass
through the point.  (We use the terms ``multiple point'' and
``intersection point'' synonymously here.)  In the column marked ``mult. pt.'',
we list information about these multiple points, including the number and the
index.  The symbol $cc$ means that the multiple points occur in complex
conjugate pairs, while ``real'' means that the multiple points occur on the
real axis in the $z$ (equivalently, $v$) plane.  For a normal double point,
i.e., a multiple point of index 2, one may assign an angle $\theta_{cr}$ as the
angle between the tangents of the two branches (arcs) of the curve which pass
through the point.  All three homopolygonal lattices have multiple
points on the curves (including line segments) across which the free
energy is non-analytic.  Each of these multiple points has index 2 and
$\theta_{cr} = \pi/2$, i.e., the two branches cross in an orthogonal manner.
Our results as shown in Figs. 5 and 6 show that
the multiple points for the $(3 \cdot 6 \cdot 3 \cdot 6)$ and
$(3 \cdot 12^2)$ lattices also have index 2 and $\theta_{cr}=\pi/2$.
Note that because the
transformation (\ref{bilinear}) relating $z$ and $v$ and also the
transformation $u=z^2$ are conformal mappings, they preserve angles, so that a
given $\theta_{cr}$ is the same for a multiple point in the $z$, $v$, and $u$
plane.  For all of these four lattices, the multiple points occur as complex
conjugate pairs as functions of $z$.  A
general observation is that the heteropolygonal lattices have more multiple
points on their curves than the homopolygonal lattices.  The
situation with the $(4 \cdot 8^2)$ lattice is the most complicated, since in
this case the loci of points where the free energy is non-analytic form
areas instead of curves.  As one can see from Fig. 7, there are a number of
multiple points through which more than one of the boundaries of these areas
pass.  There are 18 such points in all; of these, 14 consist of complex
conjugate pairs while the remaining four lie on the real $z$ or $v$ axis.
Of the 14 complex conjugate multiple points, 12 have index $n_b=2$ (with
various values of crossing angles $\theta_{cr}$).  The four multiple points
lying on the real axis (namely, $z_c = z_{1+}$, $1/z_c$, $z_{1-}$, and
$1/z_{1-}$) have index $n_b=2$ and $\theta_{cr}=0$.  In the terminology of
algebraic geometry, they are therefore (simple) tacnodes.  We recall that a
(simple) tacnode is defined as a multiple point of an algebraic curve
through which two branches pass, in an osculating manner, i.e. such that
their tangents coincide where the branches touch, so that $n_b=2$ but the
number of distinct tangents at the point is $n_t=1$ \cite{alg}.
We have denoted this in Tables 2 and 4
by the notation ``tn'', standing for ``tacnode''.
A generalised tacnode is a multiple point of an algebraic
curve for which some of the tangents of the branches passing through the
point coincide, so that $n_b > n_t$ (and $n_b$ may be greater than 2).
The $(4 \cdot 8^2)$ lattice is the first one which we have studied which
exhibits tacnodal points on the loci of points where the free energy is
non-analytic.  We come next to the multiple points at $\pm i$.  These are,
again, of a type unprecedented in any of the cases which we have previously
studied; they have index 4 and are the first example of a multiple point
of index higher than 2. Of the 4 branches which pass through this point, two
cross at an angle of $\pi/2$ with respect to each other, the northeast --
southwest and northwest -- southeast curves), while two others pass through the
points in a north -- south direction, hence with a crossing angle
of $\pi/4$ with respect to the former two curves and a crossing angle of 0 with
respect to each other.  This multiple point is therefore a higher-order
tacnodal point, with $n_b=4$ and $n_t=3$.
Finally, in the last column of Table 4 we list the number
of endpoints of the algebraic curves and in which complex-temperature phase
these endpoints lie.  For all cases, the endpoints occur in complex conjugate
pairs.

   For lattices with even $q$, viz., the square, hexagonal, and kagom\'{e}
lattices, a more compact way to present the complex-temperature phase
diagrams is in the $u$ plane.  In Table 5 we list the analogous
characteristics for these lattices.

\begin{table}
\begin{center}
\begin{tabular}{|c|c|c|c|} \hline \hline & & & \\
lattice & $N_O$ & mult. pt. & endpt. \\
& & & \\ \hline \hline
sq $(4^4)$                    & 1 & 2: 2, cc & 0 \\ \hline
tri $(3^6)$                   & 1 & 2: 2, cc & 2, cc, in FM \\ \hline
hex $(6^3)$                   & 0 & 2: 2, cc & 2, cc, in PM \\ \hline
$(3 \cdot 6 \cdot 3 \cdot 6)$ & 2 & 4: 2, cc & 4, cc, in PM \\ \hline
$(3 \cdot 12^2)$              & 2 & 4: 2, cc & 4, cc, in PM \\ \hline
$(4 \cdot 8^2)$               & 3 & 18; see below:  & 0     \\ \hline
                              &   & 12: 2, cc & \\ \hline
                              &   &  2: 3, cc, tn ($\pm i$) & \\ \hline
                              &   &  4: 2, real, tn & \\ \hline \hline
\end{tabular}
\end{center}
\caption{Comparison of some properties of complex-temperature phase diagrams,
as functions of $z$ or equivalently $v$,
for the Ising model on homopolygonal and heteropolygonal Archimedean lattices.
$N_O$ denotes the number of O phases.  In column marked ``mult. pt.'', the
entry 2: 2, cc means that there are 2 multiple points, each of index 2, and
they occur as complex conjugate pairs, etc. for other entries.  The notation
``tn'' means a tacnodal multiple point.  For the $(4 \cdot 8^2)$ lattice we
have listed the properties of the various types of multiple points underneath
the general number, 18.   See text for further discussion.}
\label{table4}
\end{table}

\begin{table}
\begin{center}
\begin{tabular}{|c|c|c|c|} \hline \hline & & & \\
lattice & $N_O$ & mult. pt. & endpt. \\
& & & \\ \hline \hline
sq $(4^4)$                    & 0 & 1: 2, real & 0 \\ \hline
tri $(3^6)$                   & 0 & 1: 2, real & 1, real, in FM \\ \hline
$(3 \cdot 6 \cdot 3 \cdot 6)$ & 1 & 2: 2, cc   & 2, cc, in PM \\ \hline
\end{tabular}
\end{center}
\caption{Comparison of some properties of complex-temperature phase diagrams,
as functions of $u$, for the Ising model on lattices with even coordination
number $q$.  Notation is as in Table 4.}
\label{table5}
\end{table}

\section{Conclusions}
\label{conclusions}

   In this paper, using exact results, we have determined the
complex-temperature phase diagrams and singularities of the spontaneous
magnetisation on three heteropolygonal Archimedean lattices,
$(3 \cdot 6 \cdot 3 \cdot 6)$ (kagom\'{e}), $(3 \cdot 12^2)$, and
$(4 \cdot 8^2)$ (bathroom tile).  This study reveals a rich variety of
complex-temperature singularities and provides an interesting comparison with
the situation for the three homopolygonal lattices.  In particular, we have
found the first example of a lattice where, even for equal spin-spin exchange
couplings, the nontrivial non-analyticities of the free energy lie in areas
rather than on curves, in the $z$ (or equivalently, $v$) plane.  We have also
given a simple explanation of why this happens.

This research was supported in part by the NSF grant PHY-93-09888.

\vfill
\eject

\vfill
\eject

\begin{center}
{\bf Figure Captions}
\end{center}

 Fig. 1. \ The $(3 \cdot 6 \cdot 3 \cdot 6)$ (kagom\'{e}) lattice.

 Fig. 2. \ The $(3 \cdot 12^2)$ lattice.

 Fig. 3. \ The $(4 \cdot 8^2)$ (bathroom tile lattice.

 Fig. 4. \ The $[4 \cdot 8^2]$ (union jack) lattice, which is the Laves lattice
dual to the Archimedean lattice $(4 \cdot 8^2)$.

 Fig. 5. \ Complex-temperature phase diagram for the Ising model on the
$3 \cdot 6 \cdot 3 \cdot 6$ (kagom\'{e}) lattice, in the variable (a) $u$
(b) $z$ (c) $v$.

 Fig. 6. \ Complex-temperature phase diagram for the Ising model on the
$3 \cdot 12^2$ lattice, in the variable (a) $z$ (b) $v$.

 Fig. 7. \ Complex-temperature phase diagram for the Ising model on the
$4 \cdot 8^2$ (bathroom tile) lattice, in the variable (a) $z$ (b) $v$.

\vfill
\eject

\end{document}